\documentclass[%
 reprint,
superscriptaddress,
amsmath,amssymb,aps,prl,floatfix,multicol]{revtex4-2}

\usepackage{bbm, bm, mathtools, hyperref, graphicx, xcolor, float} 

\begin{document}

\title[IST] {Fractional Integrable Nonlinear Soliton Equations}

\author{Mark J. Ablowitz}
\address{Department of Applied Mathematics, University of Colorado, Boulder, Colorado 80309, U.S.A.}
\thanks{}

\author{Joel B. Been}
\address{Department of Applied Mathematics and Statistics, Colorado School of Mines, Golden, Colorado 80401, U.S.A.}
\address{Department of Physics, Colorado School of Mines, Golden, Colorado 80401, U.S.A.}
\thanks{}

\author{Lincoln D. Carr}
\address{Department of Applied Mathematics and Statistics, Colorado School of Mines, Golden, Colorado 80401, U.S.A.}
\address{Department of Physics, Colorado School of Mines, Golden, Colorado 80401, U.S.A.}
\address{Quantum Engineering Program, Colorado School of Mines, Golden, Colorado 80401, U.S.A.}
\thanks{}

\begin{abstract}

Nonlinear integrable equations serve as a foundation for nonlinear dynamics, and fractional equations are well known in anomalous diffusion. We connect these two fields by presenting the discovery of a new class of integrable fractional nonlinear evolution equations describing dispersive transport in fractional media. These equations can be constructed from  nonlinear integrable equations using a widely generalizable mathematical process utilizing completeness relations, dispersion relations, and inverse scattering transform techniques. As examples, this general method is used to characterize fractional extensions to two physically relevant, pervasive integrable nonlinear equations: the Korteweg–deVries and nonlinear Schr\"odinger equations. These equations are shown to predict super-dispersive transport of non-dissipative solitons in fractional media. 
\end{abstract}

\maketitle


Fractional calculus is an effective tool when describing physical systems with power law behavior such as in anomalous diffusion, where the mean squared displacement is proportional to $t^{\alpha}$, $\alpha>0$ \cite{west_1987,random_walk,west_1997,anomalous_issue}. 
This form of transport has been observed extensively in biology \cite{saxton_2007,bronstein_2009,weigel_2011,regner_2013}, amorphous materials \cite{scher_1975,pfister_1977,gu_1996}, porous media \cite{benson_2000,benson_2001,meerschaert_2008}, and climate science \cite{koscielny_1998} amongst others. Equations in multiscale media can express fractional derivatives in any governing term \cite{west_colloquium,zhong2016spatiotemporal}, including dispersion, such as found in the 1D nonlinear Schr\"odinger equation (NLS) in optics \cite{Ablowitz2,Ablowitz2011,malomed2021optical,longhi2015fractional,li2020metastable,zeng2021bubbles,he2021propagation} and  the Korteweg-deVries equation (KdV) in water waves \cite{KDV}.  In the case of integer derivatives, NLS and KdV are famously integrable equations, leading to solitonic solutions and an infinite set of conservation laws \cite{Ablowitz1}.  Integrable equations are key signposts in nonlinear dynamics as they provide exactly solvable cases and, moreover, are an essential element of Kolmogorov-Arnold-Moser (KAM) theory underlying our understanding of chaos.  The fundamental solution of 1D dispersive integrable equations is the soliton, a robust nondispersive localized wave.  While in the space of possible nonlinear evolution equations integrable cases are extremely rare, they arise frequently in application.

In this Letter, we present a new class of integrable \emph{fractional} nonlinear evolution equations which predict super-dispersive transport in fractional media. Fractional media is ``rough'' or multiscale media that is neither regular nor random; it includes fractals but is more general as it need not be self-similar. We use the fractional NLS (fNLS) and fractional KdV (fKdV) equations as case studies. We show their integrability, demonstrate exact fractional soliton solutions, and make physical predictions about the speed of these localized 
waves. 
To date, to our knowledge, no nonlinear fractional evolution equation has been known to be integrable.

The building blocks of our demonstration are three mathematical ingredients.  Two are familiar to physicists as they are well known concepts in physics.  They are completeness and the dispersion relations.  However, in our case the dispersion relation will use fractional, rather than integer, derivatives.  The third building block is the fundamental ingredient of integrability, namely the inverse scattering transform (IST), well known to researchers in nonlinear dynamics.


Different versions of the fNLS equation have been studied in, e.g., \cite{malomed2021optical,qiu2020stabilization,li2021symmetry,al2018high}, and soliton type solutions have been found, but unlike the fNLS and fKdV equations that we introduce, none of these are integrable. The fractional operators in the fNLS and fKdV equations are nonlinear generalizations of the Riesz fractional derivative. In fact, the linear limit of the fNLS equation is the well known fractional Schr\"odinger equation derived using a Feynman path integral over L\'evy flights \cite{laskin2000fractional,laskin2018}. Fractional equations defined using the Riesz fractional derivative (alternately termed the Riesz transform \cite{riesz} or fractional Laplacian \cite{what_is_frac_lap}) are effective tools when describing behavior in complex systems because the Riesz fractional derivative is closely related to non-Gaussian statistics \cite{frac_stoch}. It has found physical applications in describing movement of water in porous media \cite{frac_porous}, transport of temperature in fluid dynamics \cite{quasi_geo}, and power law attenuation in materials \cite{power_law_attn} amongst many others \cite{frac_lap,diff_appl_1,diff_appl_2}. 
 
The KdV and NLS equations arise in many physical problems. The KdV equation is applicable in shallow water waves, internal waves, fluid dynamics, plasma physics, and lattice dynamics amongst others \cite{KDV}. Furthermore, KdV is a universally important equation whenever weak dispersion balances weak quadratic nonlinearity cf. \cite{Ablowitz2,Ablowitz2011}. Similarly, the NLS equation arises in the quasi-monochromatic approximation with dispersion balancing weak nonlinearity and occurs widely in physical applications, e.g. water waves, nonlinear optics, spin waves in ferromagnetic films, plasma physics, Bose-Einstein condensates, etc.~\cite{Ablowitz2,Ablowitz2011,bronski_2001,boardman_1994}. 
The KdV equation was shown to be solvable using the IST and to admit soliton solutions when associated with the  linear time-independent Schr\"odinger equation  in \cite{KDV1967}. Then, the NLS equation with decaying data was solved and shown to possess solitons via the IST in \cite{ZS72}. Soon after, the method was extended to the modified KdV and  sine-Gordon equations as well as general classes of equations written in terms of a linearized dispersion relation \cite{AKNS,Ablowitz2011}. IST is now a large field cf. \cite{Ablowitz2,Ablowitz1,calogero1982,Novikov1984,Remoissinet2013}.

Here we show  how  to extend this formulation to encompass fractional integrable nonlinear evolution equations. As examples of this technique, we show that fKdV and fNLS are solvable by the IST. These are two examples of many possible fractional integrable equations that can be characterized by this method.

{\it The IST and anomalous dispersion relations} ---
It is well known that linear evolution equations for $q = q(x,t)$ of the form
\begin{align}
    q_{t} + \gamma(\partial_{x}) q_{x} = 0, \label{eqn:linear_wave_equation}
\end{align}
can be solved by Fourier transforms when $\gamma(\partial_{x})$ is a rational function of $\partial_{x}$; cf. \cite{Ablowitz2011}. We can do this because the completeness of plane waves gives an integral representation of $\gamma(\partial_{x})$.
%
 The solution to (\ref{eqn:linear_wave_equation}) is explicitly
\begin{align}
    q(x,t) = \frac{1}{2 \pi} \int_{-\infty}^{\infty} dk \hat{q}(k,0) e^{i k x - i k \gamma(i k) t}, \label{eqn:linear_wave_equation_solution}
\end{align}
where $\hat{q}(k,0)$ is the Fourier transform of $q(x,t)$ with respect to $x$ evaluated at $t = 0$. However, as Riesz showed \cite{riesz}, the solution (\ref{eqn:linear_wave_equation_solution}) makes sense for much more general $\gamma$. Specifically, Fourier Transforms can be used to solve linear fractional evolution equations, e.g., $\gamma(\partial_{x}) = |-\partial_{x}^2|^{\epsilon}$ with $2\epsilon$ the order of the fractional derivative; we take $0<\epsilon<1$ throughout this letter.

Here we show that similar analysis applies to nonlinear evolution equations using the IST. We do this by associating a class of integrable nonlinear equations with a linear scattering problem (ingredient 1, IST), characterizing the fractional equation with an anomalous dispersion relation (ingredient 2, dispersion), and defining the fractional operator associated with this dispersion relation using the completeness of squared eigenfunctions of the scattering equation (ingredient 3, completeness). 

We will apply ingredients $1$ and $2$ to find the fKdV and fNLS equations, and use ingredient $3$ to define the fractional operators in these equations. Associated with the non-dimensionalized time-independent Schr\"odinger equation for $v(x,t)$ with potential $q(x,t)$
\begin{equation}
    v_{xx} + \left( k^2 + q(x,t) \right) v = 0 , ~~|x|<\infty,
\label{eqn:schrodinger_scattering}
\end{equation}
is the following class of integrable nonlinear equations for $q(x,t)$ \cite{AKNS}
\begin{equation}
    q_{t} + \gamma(L^{A}) q_{x} = 0, \quad   L^{A} \equiv -\frac{1}{4} \partial_{x}^2 - q + \frac{1}{2} q_{x} \int_{x}^{\infty} dy.  
    \label{eqn:general_evolution_equation}
\end{equation}
where $\int_{x}^{\infty}dy$ operates on the function to which $L^A$ is applied by integrating it. Hence, equation (\ref{eqn:general_evolution_equation}) can be solved by the IST using (\ref{eqn:schrodinger_scattering}). We obtain the fKdV equation by choosing $\gamma(L^{A}) = -4 L^{A} \left|4 L^{A}\right|^{\epsilon}$; this will be justified shortly.

Similarly, associated with the following $2\times2$ scattering problem --- termed the Ablowitz-Kaup-Newell-Segur (AKNS) system --- for the vector-valued function $\mathbf{v}(x,t) = \left(v_{1}(x,t),v_{2}(x,t)\right)^{T}$ ($T$ represents transpose) 
\begin{align}
    v^{(1)}_{x} &= - i k v^{(1)} + q(x,t) v^{(2)}, \label{eqn:AKNS_scattering_1} \\
    v^{(2)}_{x} &= i k v^{(2)} + r(x,t) v^{(1)}, \label{eqn:AKNS_scattering_2}
\end{align}
is the set of integrable nonlinear equations \cite{AKNS}
\begin{align}
    \sigma_{3}\partial_t\mathbf{u} + 2 A_{0}(\mathbf{L}^{A}) \mathbf{u} = 0, \quad \sigma_{3} = \begin{pmatrix} 1 & 0 \\ 0 & -1 \end{pmatrix},
    \label{eqn:AKNS_general_evolution_equation}
\end{align}
where $\mathbf{u} = \left( r, q \right)^{T}$ and the operator 
\begin{align}
    \mathbf{L}^{A} \equiv \frac{1}{2 i} \begin{pmatrix} \partial_{x} - 2 r I_{-} q & 2 r I_{-} r \\
    - 2 q I_{-} q & -\partial_{x} + 2 q I_{-} r \end{pmatrix} \label{eqn:AKNS_L_operator} 
\end{align}
with $I_{-} = \int_{-\infty}^{x} dy$. Note that $I_{-}$ operates both on the function immediately to its right and the functions to which $\mathbf{L}^{A}$ is applied. Taking  $r = \mp q^{*}$, $*$ the complex conjugate, and $A_{0}(\mathbf{L}^{A}) = 2 i (\mathbf{L}^{A})^2 |2 \mathbf{L}^{A}|^{\epsilon}$ we find fNLS to be the second component of (\ref{eqn:AKNS_general_evolution_equation}).

These definitions are justified when we note that $\gamma(L^{A})$ and $A_{0}(\mathbf{L}^{A})$ can be related to the dispersion relation of the linearization of (\ref{eqn:general_evolution_equation}) and (\ref{eqn:AKNS_general_evolution_equation}). Specifically, if we put $q = e^{i (k x - w(k) t)}$  into the linearizations of (\ref{eqn:general_evolution_equation}) and (\ref{eqn:AKNS_general_evolution_equation}), we have
\begin{align}
    \gamma(k^2) = \frac{w_{K}(2 k)}{2 k}, \quad A_{0}(k) = -\frac{i}{2} w_{S}(-2k), \label{eqn:dispersion_relations}
\end{align}
where $w_K$ is the dispersion relation for the linear fKdV equation and $w_S$ is the same for the linear fractional Schr\"odinger equation. Therefore, $\gamma(L^{A})$ and $A_{0}(\mathbf{L}^{A})$ for fKdV and fNLS are generated from the dispersion relations for linear fKdV and the linear fractional Schr\"odinger equation. These equations are, naturally,
\begin{align}
    q_{t} + \left|- \partial_{x}^{2}\right|^{\epsilon} q_{xxx} = 0, \quad i q_{t} = \left|-\partial_{x}^2\right|^{\epsilon/2} q_{xx}, \label{eqn:linear_fractional_KdV}
\end{align}
where $|- \partial_{x}^{2}|^{\epsilon}$ is the Riesz fractional derivative. So, the corresponding dispersion relations are $w_{K}(k) = - k^3 |k|^{2 \epsilon}$ and $w_{S}(k) = -k^2 |k|^{\epsilon}$ which lead to the aforementioned definitions of $\gamma(L^{A})$ and $A_{0}(\mathbf{L}^{A})$.

{\it Spectral definitions of fKdV and fNLS by completeness} ---
To define the fKdV and fNLS equations we need to determine what operating on a function with $\gamma(L^{A})$ or $A_{0}(\mathbf{L}^{A})$ means. We do this using ingredient 3, completeness of the associated linear scattering system.


In \cite{AKNS} it was shown that the eigenfunctions of $L^{A}$ are any of the three functions: $\{\partial_x \varphi^2, \partial_x \psi^2, \partial_x (\varphi \psi)\}$ which we represent generically as $\Psi^{A}$, each with eigenvalue $\lambda = k^2$. Here $\psi$ and $\varphi$ solve the time-independent Schrodinger equation (\ref{eqn:schrodinger_scattering}) subject to appropriate asymptotic boundary conditions at $x=\pm \infty$. Furthermore, the eigenfunctions of $\mathbf{L}^{A}$ are $\mathbf{\Psi}^{A}$ and $\overline{\mathbf{\Psi}}^{A}$ each with eigenvalue $\lambda = k$. These may be written in terms of solutions to equations (\ref{eqn:AKNS_scattering_1}) and (\ref{eqn:AKNS_scattering_2}) (see Supplemental Material \cite{supplemental_material}).

Starting from $\gamma(L^{A})$ and $A_{0}(\mathbf{L}^{A})$ operating on $\Psi^{A}$ and $\mathbf{\Psi}^{A}$, we can write
\begin{align}
    \gamma(L^{A}) \Psi^{A} &= \gamma(k^2)  \Psi^{A}, \label{eqn:KdV_eigenfunction_definition} \\
    A_{0}(\mathbf{L}^{A}) \mathbf{\Psi}^{A} &= A_{0}(k) \mathbf{\Psi}^{A}.
    \label{eqn:NLS_eigenfunction_definition}
\end{align}
To extend this to $\gamma(L^{A})$ and $A_{0}(\mathbf{L}^{A})$ operating on any function, we need to be able to express any function in terms of $\Psi^{A}$ and $\mathbf{\Psi}^{A}$, i.e. we need a completeness relation for each set of eigenfunctions.


In \cite{sachs} it was shown that the eigenfunctions $\Psi^{A}$ are complete. Assuming $q(x,t)$ is sufficiently decaying and smooth in $x$, an arbitrary, and sufficiently regular, function $h(x)$ may be expanded in terms of the eigenfunctions $\Psi^A$ as 
 \begin{align}
    h(x) = \int_{\Gamma_{\infty}} dk ~\frac{\tau^2(k)}{4 \pi i k} \int_{-\infty}^{\infty} dy ~G(x,y,k) h(y),
    \label{eqn:square_schrodinger_completness}
\end{align}
where time is suppressed and $\Gamma_{\infty}= \lim_{R \to \infty} \Gamma_{R}$ with $\Gamma_{R}$ the semicircular contour in the upper half plane evaluated from $k = - R$ to $k = R$. $\tau$ is the transmission coefficient defined by the relation $\varphi(x,k) \tau(k) = \psi(x,-k) + \rho(k) \psi(x,k)$, $\rho$ is the reflection coefficient, and
\begin{align}
    G(x,y,k) = \partial_{x} (\psi^2(x,k) \varphi^2(y,k) - \varphi^2(x,k) \psi^2(y,k)). \label{eqn:square_schrodinger_completness_kernel}
\end{align}
This completeness relation reduces to Fourier completeness in the linear limit. From (\ref{eqn:KdV_eigenfunction_definition}) and (\ref{eqn:square_schrodinger_completness}) the operation of $\gamma(L^{A})$ on a sufficiently smooth and decaying function $h$ follows as
\begin{align}
 \gamma(L^{A}) h(x) &= \int_{\Gamma_{\infty}}\! dk\gamma(k^2)\frac{\tau^2(k)}{4 \pi i k}\int_{-\infty}^{\infty}\! dy G(x,y,k) h(y).
 \label{eqn:gamma_L_spectral_expansion}
\end{align}
Hence, equations (\ref{eqn:square_schrodinger_completness})-(\ref{eqn:gamma_L_spectral_expansion}) provide an explicit representation of fKdV, i.e. equation (\ref{eqn:general_evolution_equation}) with $\gamma(L^{A}) = - 4 L^{A} \left| 4 L^{A} \right|^{\epsilon}$, which may be written as
\begin{align}
    q_{t} +\! \int_{\Gamma_{\infty}}\!\! dk |4 k^2|^{\epsilon}  \frac{\tau^2(k)}{4 \pi i k}\int_{-\infty}^{\infty}\! dy G(x,y,k) \left( 6 q q_{y} + q_{yyy} \right)\! =\! 0. \label{eqn:fKdV}
\end{align}
Notice that equation (\ref{eqn:fKdV}) is in non-dimensional coordinates $x$ and $t$. In the linear limit $q\to0$, we have $\gamma(L^{A}) \to \gamma\left(-\partial_{x}^2/4\right)$. So, for fKdV, $\gamma(L^{A}) \to -\partial_{x}^{2} \left|- \partial_{x}^{2}\right|^{\epsilon}$ which is the Riesz fractional derivative. If  we then set $\epsilon = 0$, we recover the KdV equation:
\begin{align}
    q_{t} + 6 q q_{x} + q_{xxx} = 0. \label{eqn:KdV}
\end{align}
We note that $\tau(k,t)$ has a finite number of simple poles along the imaginary axis denoted $k_{j} = i \kappa_{j}$ for $j = 1, 2, ..., J$, so the above representation can be evaluated by contour integration (see Supplemental Material \cite{supplemental_material}). 

Similarly, the eigenfunctions $\mathbf{\Psi}^{A}$ are also complete \cite{Kaup76}. Thus, we can write the operation of $A_{0}(\mathbf{L}^{A})$ on a sufficiently smooth and decaying vector-valued function $\mathbf{h}(x) = \left( h_{1}(x),h_{2}(x) \right)^{T}$ as
\begin{align}
    A_{0}(\mathbf{L}^{A}) \mathbf{h}(x) \! &= \!\!\sum_{n = 1}^{2}\!\int_{\Gamma_{\infty}^{(n)}}\! \!\! dk  A_{0}(k) f_{n}(k) \!\! \int_{-\infty}^{\infty}\! dy\mathbf{G}_{n}(x,\!y,\!k)\mathbf{h}(y),  \label{eqn:A0_L_spectral_expansion} \\
    \mathbf{G}_{1}(x,y,k) &= \mathbf{\Psi}^{A}(x,k) \mathbf{\Psi}(y,k)^{T}, \quad f_{1}(k) = -\tau^2(k)/\pi, \notag \\
    \mathbf{G}_{2}(x,y,k) &= \overline{\mathbf{\Psi}}^{A}(x,k) \overline{\mathbf{\Psi}}(y,k)^{T}, \quad f_{2}(k) = \overline{\tau}^2(k)/\pi, \notag
 \end{align}
where $\Gamma_{R}^{(1)}$ ($\Gamma_{R}^{(2)}$) is the semicircular contour in the upper (lower) half plane evaluated from $-R$ to $+R$; $\mathbf{\Psi}(x,k)$, $\overline{\mathbf{\Psi}}(x,k)$ are eigenfunctions of $\mathbf{L}$; $\mathbf{\Psi}^{A}(x,k)$, $\overline{\mathbf{\Psi}}^{A}(x,k)$ are eigenfunctions of $\mathbf{L}^{A}$; and $\tau(k)$, $\overline{\tau}(k)$ are transmission coefficients defined similarly to fKdV. Notice that $\mathbf{G}_{n}$ are $2\times2$ matrices (see Supplemental Material \cite{supplemental_material}).

Thus equation (\ref{eqn:A0_L_spectral_expansion}) gives a representation for the fNLS, equation (\ref{eqn:AKNS_general_evolution_equation}) with $A_{0}(\mathbf{L}^{A}) = 2 i (\mathbf{L}^{A})^2 |2 \mathbf{L}^{A}|^{\epsilon}$ and $r = \mp q^{*}$; see the Supplemental Material \cite{supplemental_material}. 
In the linear limit, fNLS is represented in terms of the Riesz fractional derivative and for $\epsilon = 0$ we recover NLS:
\begin{align}
    i q_{t} = q_{xx} \pm 2 q^2 q^{*}. \label{eqn:NLS}
\end{align}
With explicit expressions for $\gamma(L^{A})$ and $A_{0}(\mathbf{L}^{A})$ in equations (\ref{eqn:gamma_L_spectral_expansion}) and (\ref{eqn:A0_L_spectral_expansion}), the fKdV and fNLS equations are characterized. Further, because these equations are inside of the time-independent Schr\"odinger and AKNS classes of integrable nonlinear equations in (\ref{eqn:general_evolution_equation}) and (\ref{eqn:AKNS_general_evolution_equation}), fKdV and fNLS are solvable by the IST.

{\it Soliton solutions of fKdV and fNLS} --- Given an initial state $q(x,0)$ with sufficient smoothness and decay, we can solve fKdV and fNLS, i.e. obtain $q(x,t)$, using the IST. To do this, we first map the initial state into scattering space, evolve the resulting scattering data in time, and reconstruct the solution in physical space from these data. It turns out that solving fKdV and fNLS are remarkably similar to solving KdV and NLS. 

We note that, given the explicit representation of fKdV in equation (\ref{eqn:fKdV}), and fNLS in Supplemental Material \cite{supplemental_material}, these equations can also be solved numerically in discrete time by finding the kernels $G/\mathbf{G}_{j}$ and evaluating the integrals with respect to $y$ and $k$ at each time step.

The fractional soliton solutions of fKdV and fNLS are given in equations (\ref{eqn:KdV_one_soliton}-\ref{eqn:NLS_one_soliton}). These correspond to reflectionless bound states of the Schr\"odinger and AKNS scattering problems with one complex eigenvalue $k_{K} = i \kappa$ and $k_{S} = \xi + i \eta$ respectively:
\begin{align}
    q_{K}(x,t) &= 2 \kappa^{2} \text{sech}^{2}{(\kappa [(x - x_{1}) -(4 \kappa^2)^{1+\epsilon} t])}, \label{eqn:KdV_one_soliton}\\
    q_{S}(x,t) &= 2 \eta e^{- 2 i \xi x + 4 i ( \xi^2 - \eta^2 ) |2 k_{S}|^\epsilon t} \label{eqn:NLS_one_soliton}  \text{sech}{(z_{\epsilon}(x,t))},
\end{align}
where $z_{\epsilon}(x,t) = 2 \eta( x - x_{0} - 4 \xi |2 k_{S}|^\epsilon t )$ and $x_0$, $x_1$ can be characterized in terms of scattering data.

It can also be shown that the fractional solitons solve their respective equations by evaluating $\gamma(L^{A}) \partial_{x}q_{K}$ and $A_{0}(\mathbf{L}^{A}) \partial_{x}q_{S}$ using contour integration methods (this computation for the fKdV equation is given in the Supplemental Material \cite{supplemental_material}.) Further, higher order solitons can be calculated and their interactions are elastic.


\textit{Physical Predictions} -- The fKdV and the fNLS equations describe the transport of fluid and photons in multiscale fluid channels and laser fiberoptic systems, respectively. The multiscale characteristic of these materials represents a certain ``roughness'' which is averaged over in fKdV and fNLS. The solitonic solutions of these equations describe how localized waves of fluid/probability are transported in such systems.
Both fKdV and fNLS predict solitons with anomalous motion, that is, super-dispersive transport where speeds are larger than expected from regular or ordered systems (note that sub-dispersive transport can also be realized by modifying the dispersion relation). Specifically, the group velocity of fKdV and fNLS and the phase velocity of fNLS are
\begin{align}
    v_{K}(\epsilon,\kappa) &= \left(4 \kappa^2\right)^{1+\epsilon} \label{eqn:KdV_velocity} \\
    v_{S}(\xi,\eta) &= 2^{2+\epsilon} \xi (\xi^2 + \eta^2)^{\epsilon/2} \label{eqn:NLS_velocity} \\
    v_{\theta}(\xi,\eta) &= 2^{1+\epsilon} \left(\xi^2 - \eta^2\right) \left( \xi^2 + \eta^2 \right)^{\epsilon/2}/\xi \label{eqn:NLS_phase_velocity}
\end{align}
%
In a wave tank of height $5$ cm we expect solitons with amplitude and KdV speed around $2/3$ cm and $0.3$ cm/s, respectively. One can similarly associate physical values to solitons in fiberoptics \cite{mollenauer1980}, spin waves in ferromagnetic films \cite{ustinov_2010}, Bose-Einstein condensates \cite{Strecker_2003}, or any of the many other contexts in which NLS is applicable.

\begin{figure}[H]
\begin{centering}
\includegraphics[width=0.35\textwidth]{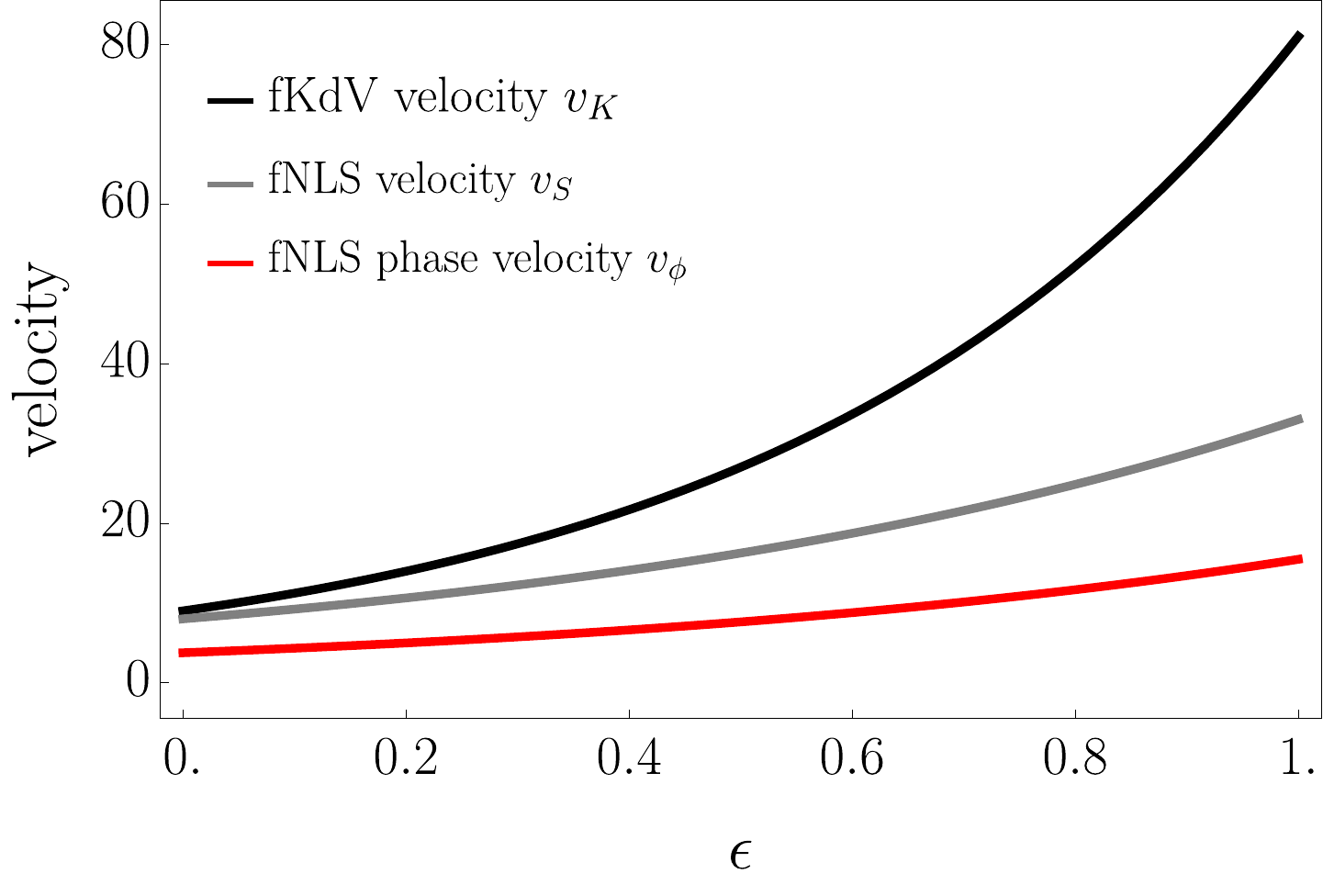}
\caption{\label{fig:velocity_plots}
    Localized waves predicted by the fKdV and fNLS equations, (\ref{eqn:KdV_velocity}-\ref{eqn:NLS_phase_velocity}), show super-dispersive transport as their velocity increases as $\epsilon$ increases from $0$ to $1$. Like anomalous diffusion where the mean squared displacement is proportional to $t^{\alpha}$, the velocity in anomalous dispersion is proportional to $A^{\epsilon}$, where $A$ is the amplitude of the wave. The parameter values used are $\kappa = 3/2$, $\xi = 2$, and $\eta = 1/2$.
}
\end{centering}
\end{figure}

Figure \ref{fig:velocity_plots} shows the velocities in equations (\ref{eqn:KdV_velocity}-\ref{eqn:NLS_phase_velocity}) as they interpolate between KdV (NLS) for $\epsilon = 0$ and $\epsilon = 1$.  
Notice that fKdV and fNLS predict a power law relationship between the amplitude of the wave, $\kappa^2$ and $\eta$ respectively, and the speed of the wave characterized by $\epsilon$. Experimentally verifying these relations relies on comparing the amplitude of water waves and the amplitude and phase of laser pulses in optical fibers to their speed in multiscale media. 

\begin{figure}[H]
\begin{centering}
\includegraphics[width=0.35\textwidth]{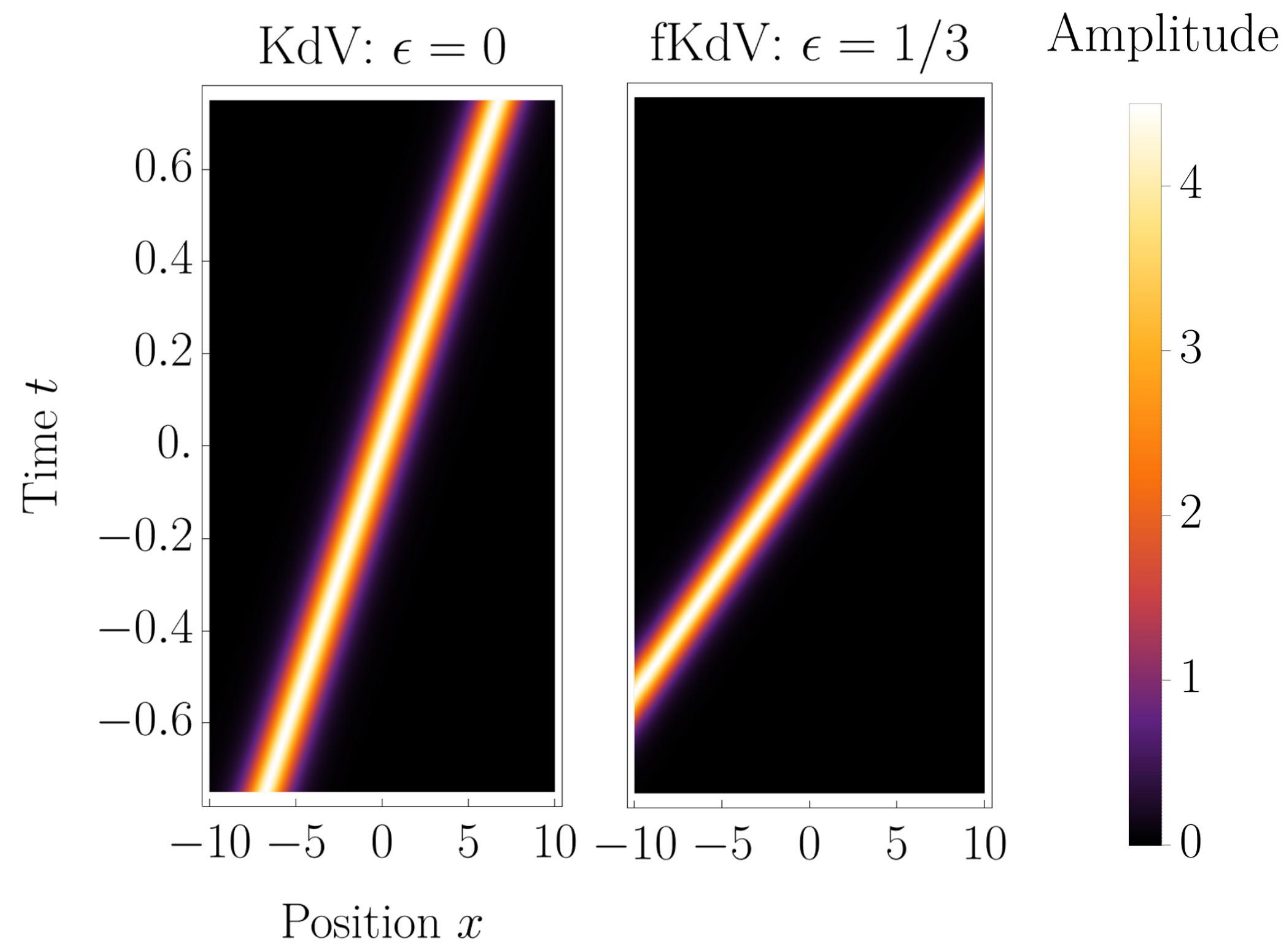}
\caption{\label{fig:fKdV_solitons}
    Note that soliton solutions to the fKdV equation propagate without dissipating or spreading out. The parameter values used are $\kappa = 3/2$ and $x_{0} = 0$.
}
\end{centering}
\end{figure}

Importantly, the physical properties of fractional solitons, besides the change in velocity described by equations (\ref{eqn:KdV_velocity}-\ref{eqn:NLS_phase_velocity}), are identical to regular ones. From figure \ref{fig:fKdV_solitons}, fractional solitons propagate without dissipating or spreading out. An open question is to compare the solitons predicted by fKdV and fNLS to solitary waves predicted by other, non-integrable versions of these equations. This could be done by studying how the velocity of each equation varies with the fractional parameter $\epsilon$ and whether soliton-soliton interactions are elastic or inelastic and what the predicted phase shifts are.

\textit{Conclusion} --- We have demonstrated a new class of integrable equations, namely 1D fractional integrable nonlinear evolution equations, derivable from a general method.  As ubiquitous examples of this class we presented integrability and solitonic solutions of the fractional nonlinear Schr\"odinger and Korteweg-deVries equations. We demonstrated the three basic mathematical ingredients of our procedure: completeness, dispersion relations, and inverse scattering transform techniques. We also gave fractional soliton solutions to these equations and demonstrated super-dispersive transport as a physical implication of the equations. Such fractional equations model multiscale materials and open new directions in integrable nonlinear dynamics for such systems, both artificial and naturally occurring. Our method provides a context for the discovery and understanding of 1D fractional nonlinear evolution equations generally, with integrability acting as a key signpost for fractional nonlinear dynamics.

\section*{Acknowledgements}
We thank U. Al-Khawaja, A. Gladkina, J. Lewis, and M. Wu for useful discussions. This project was partially supported by NSF under grants DMS-2005343 and DMR-2002980.

\newpage

\section*{Supplemental Material}

\subsection{Scattering theory}

The inverse scattering transformation (IST) solves nonlinear wave equations by associating them with linear eigenvalue problems. The fKdV equation is associated with the time-independent Schr\"odinger equation, (\ref{eqn:schrodinger_scattering_SM}), while the fNLS equation is associated with the AKNS system, (\ref{eqn:AKNS_scattering_1_SM})-(\ref{eqn:AKNS_scattering_2_SM}). Here, we define the eigenfunctions and scattering data of the linear eigenvalue problem and provide some important properties.

\subsubsection{Scattering theory of the time-independent Schr\"odinger equation} 

The Schr\"odinger scattering problem is
\begin{equation}
    v_{xx} + \left( k^2 + q(x,t) \right) v = 0 , ~~|x|<\infty
\label{eqn:schrodinger_scattering_SM}
\end{equation}
where $v$ is the eigenfunction and $k^2$ is the eigenvalue. This is associated to the following class of integrable nonlinear equations for $q(x,t)$
\begin{equation}
    q_{t} + \gamma(L^{A}) q_{x} = 0, \quad   L^{A} \equiv -\frac{1}{4} \partial_{x}^2 - q + \frac{1}{2} q_{x} \int_{x}^{\infty} dy 
    \label{eqn:general_evolution_equation_SM}
\end{equation}
where $\gamma(L^A)$ is defined in Eq. (\ref{eqn:gamma_L_spectral_expansion_SM}). Appropriate eigenfunctions are solutions to Eq. (\ref{eqn:schrodinger_scattering_SM}) with asymptotic boundary conditions
\begin{alignat}{2}
    \varphi(x,k,t) &\sim e^{- i k x}, \quad &&\overline{\varphi}(x,k,t) \sim e^{i k x}, ~~x \to -\infty, \label{eqn:asymptotic_phi} \\
        \psi(x,k,t) &\sim e^{i k x}, &&\overline{\psi}(x;k,t) \sim e^{-i k x}, ~~x \to \infty. \label{eqn:asymptotic_psi}
\end{alignat}
Because $\psi$ and $\overline{\psi}$ are linearly independent and Eq. (\ref{eqn:schrodinger_scattering_SM}) is second order, we have
\begin{align}
    \varphi(x,k,t) &= a(k,t) \overline{\psi}(x,k,t) + b(k,t) \psi(x,k,t). \label{eqn:scattering_data_phi} 
\end{align}
The eigenfunctions are related via
\begin{equation}
    \varphi(x,k,t) = \overline{\varphi}(x,-k,t), ~\psi(x,k,t) = \overline{\psi}(x,-k,t)   
    \label{eqn:conjugate_relations_psi}
\end{equation}
and the scattering data are obtained from
\begin{align}
    a(k,t) = \frac{W(\varphi, \psi)}{2 i k}, \quad b(k,t) = \frac{W(\overline{\psi}, \varphi)}{2 i k} \label{eqn:wronskian_relations}
\end{align}
where $W(u,v)$ is the Wronskian $W(u,v) = u v_{x} - v u_{x}$. 
The associated transmission and reflection coefficients of the Schr\"odinger scattering problem (\ref{eqn:schrodinger_scattering_SM}) are written in terms of $a,b$ as
\begin{align}
    \tau(k,t) = \frac{1}{a(k,t)}, \quad \rho(k,t) = \frac{b(k,t)}{a(k,t)}. \label{eqn:transmisson_and_reflection_coefficients}
\end{align}
Discrete eigenvalues correspond to zeros of $a$ at $k_{j}$, $j=1,2,...,J$ where $J$ gives the number of solitons in the solution. When $q$ is real the eigenvalues are purely imaginary; i.e. $k_j=i \kappa_j$, $\kappa_{j}$ real,  $\kappa_j>0$. At these eigenvalues, which are simple, the eigenfunctions decay exponentially --- they are bound states. At these discrete eigenvalues, the eigenfunctions are related by
\begin{align}
    \varphi_{j}(x,t)=b_j(t) \psi_{j}(x,t) \label{eqn:normalization_constant}
\end{align}
where $\varphi_{j}(x,t) = \varphi(x, k_{j},t)$, $\psi_{j}(x,t) = \psi(x,k_{j},t)$.

\subsubsection{Scattering theory of the AKNS system}

The AKNS scattering problem is:
\begin{align}
    v^{(1)}_{x} &=- i k v^{(1)} + q(x,t) v^{(2)}, \label{eqn:AKNS_scattering_1_SM} \\
    v^{(2)}_{x} &= i k v^{(2)} + r(x,t) v^{(1)} \label{eqn:AKNS_scattering_2_SM}
\end{align}
where $v^{(n)}$ represents the $n$th component of the vector $\underline{v} = [v^{(1)}, v^{(2)}]^{T}$. This is associated to the following set of integrable nonlinear equations
\begin{align}
    \sigma_{3}\partial_t\mathbf{u} + 2 A_{0}(\mathbf{L}^{A}) \mathbf{u} = 0, \quad \sigma_{3} = \begin{pmatrix} 1 & 0 \\ 0 & -1 \end{pmatrix}
    \label{eqn:AKNS_general_evolution_equation_SM}
\end{align}
where $\mathbf{u} = \left[ r, q \right]^{T}$ and the operator 
\begin{align}
    \mathbf{L}^{A} \equiv \frac{1}{2 i} \begin{pmatrix} \partial_{x} - 2 r I_{-} q & 2 r I_{-} r \\
    - 2 q I_{-} q & -\partial_{x} + 2 q I_{-} r \end{pmatrix} \label{eqn:AKNS_L_operator_SM} 
\end{align}
with $I_{-} = \int_{-\infty}^{x} dy$. With sufficient decay and smoothness, we define eigenfunctions for the AKNS system as solutions to Eqs. (\ref{eqn:AKNS_scattering_1_SM})-(\ref{eqn:AKNS_scattering_2_SM}) satisfying the boundary conditions
\begin{alignat}{2}
    \boldsymbol{\phi}(x,k,t) &\sim \begin{pmatrix}
    1 \\
    0
    \end{pmatrix} e^{-i k x},~&&\overline{\boldsymbol{\phi}}(x,k,t) \sim \begin{pmatrix}
    0 \\
    1
    \end{pmatrix} e^{i k x}, x\to-\infty, \label{eqn:AKNS_asymptotic_phi} \\
    \boldsymbol{\psi}(x,k,t) &\sim \begin{pmatrix}
    0 \\
    1
    \end{pmatrix} e^{i k x},~&&\overline{\boldsymbol{\psi}}(x,k,t) \sim \begin{pmatrix}
    1 \\
    0
    \end{pmatrix} e^{-i k x}, x\to\infty. \label{eqn:AKNS_asymptotic_psi}
\end{alignat}
The eigenfunctions $\boldsymbol{\psi}$, $\overline{\boldsymbol{\psi}}$ are linearly independent so that
\begin{align}
    \boldsymbol{\phi}(x,k,t) &= b(k,t) \boldsymbol{\psi}(x,k,t) + a(k,t) \overline{\boldsymbol{\psi}}(x,k,t), \label{eqn:AKNS_scattering_data_phi} \\
    \overline{\boldsymbol{\phi}}(x,k,t) &= \overline{a}(k,t) \boldsymbol{\psi}(x,k,t) + \overline{b}(k,t) \overline{\boldsymbol{\psi}}(x,k,t). \label{eqn:AKNS_scattering_data_psi}
\end{align}
The scattering data is obtained from
\begin{alignat}{2}
    a(k) &= W(\boldsymbol{\phi}, \boldsymbol{\psi}), \quad \overline{a}(k) &&= W(\overline{\boldsymbol{\psi}},\overline{\boldsymbol{\phi}}), \label{eqn:AKNS_wronskian_relations_a} \\
    b(k) &= W(\overline{\boldsymbol{\psi}},\boldsymbol{\phi}), \quad \overline{b}(k) &&= W(\overline{\boldsymbol{\phi}},\boldsymbol{\psi}) \label{eqn:AKNS_wronskian_relations_b}
\end{alignat}
with the Wronskian  given by $W(\mathbf{u},\mathbf{v}) = u^{(1)} v^{(2)} - u^{(2)}v^{(1)}$. The transmission and reflection coefficients, $\tau$, $\overline{\tau}$, $\rho$, and $\overline{\rho}$, are defined analogously to Eq. (\ref{eqn:transmisson_and_reflection_coefficients}).

The zeros of $a$ and $\overline{a}$ at $k_{j} = \xi_{j} + i \eta_{j}$, $\eta_{j} > 0$, $j = 1, 2, ..., J$ and $\overline{k}_{j} = \overline{\xi}_{j} + i \overline{\eta}_{j}$, $\overline{\eta}_{j} < 0$, $j = 1, 2, ..., \overline{J}$  are the eigenvalues. We assume the eigenvalues are `proper': i.e. they are simple, not on the real $k$ axis, and $J= \overline{J}$ ; cf. Ref. \cite{Ablowitz3}.
The bound state eigenfunctions are related by
\begin{align}
    \boldsymbol{\phi}_{j}(x,t) = b_{j}(t) \boldsymbol{\psi}_{j}(x,t), \quad \overline{\boldsymbol{\phi}}_{j}(x,t) = \overline{b}_{j}(t) \overline{\boldsymbol{\psi}}_{j}(x,t).
\end{align}
When $r = \mp q^{*}$, we have the symmetry reductions
\begin{align}
    \overline{\boldsymbol{\psi}}(x,k,t) &= \Sigma \boldsymbol{\psi}^{*}(x,k^{*},t), ~~ \overline{\boldsymbol{\phi}}(x,k,t) = \mp \Sigma \boldsymbol{\phi}^{*}(x,k^{*},t) \label{eqn:ANKS_conjugate_relation}
\end{align}
for the eigenfunctions where $\Sigma = \begin{pmatrix} 0 & 1 \\ \pm 1 & 0 \end{pmatrix}$. We also have 
$\overline{a}(k) = a^{*}(k^{*})$ and $\overline{b}(k) = \mp b^{*}(k^{*})$ for the scattering data.

From the scattering eigenfunctions $\boldsymbol{\psi}$ and $\boldsymbol{\phi}$, we can construct the eigenfunctions of the operator $\mathbf{L}$, $\mathbf{\Psi}$ and $\overline{\mathbf{\Psi}}$, and its adjoint $\mathbf{L}^{A}$, $\mathbf{\Psi}^{A}$ and $\overline{\mathbf{\Psi}}^{A}$
\begin{alignat}{2}
    \mathbf{\Psi}(x,k,t) &= \left[(\psi^{(1)}(x,k,t))^2, (\psi^{(2)}(x,k,t))^2 \right]^{T}, \\ ~ \overline{\mathbf{\Psi}}(x,k,t) &= \left[(\overline{\psi}^{(1)}(x,k,t))^2, (\overline{\psi}^{(2)}(x,k,t))^2 \right]^{T},  \\
    \mathbf{\Psi}^{A}(x,k,t) &= \left[(\phi^{(2)}(x,k,t))^2, -(\phi^{(1)}(x,k,t))^2 \right]^{T}, \\ \overline{\mathbf{\Psi}}^{A}(x,k,t) &= \left[(\overline{\phi}^{(2)}(x,k,t))^2, -(\overline{\phi}^{(1)}(x,k,t))^2 \right]^{T}.
\end{alignat}
We use these functions to define the fNLS equation.

\subsection{Direct Scattering}

To solve fKdV and fNLS by the inverse scattering transform, we first map the initial condition into scattering space; this is analogous to taking the Fourier transform of a linear PDE. This process involves analyzing linear integral equations for the eigenfunctions, determining their analytic properties,  and then obtaining the scattering data using Wronskian relations.

\subsubsection{Direct scattering for the time-independent Schr\"odinger equation}

The eigenfunctions $\varphi$ and $\psi$ of the time-independent Schr\"odinger Eq. (\ref{eqn:schrodinger_scattering_SM})  solve linear integral equations which have uniformly convergent Neumann series for $q(x,0)$ in $L_{2}^{1}$ \cite{Ablowitz2011}. This series can be used to construct the eigenfunctions explicitly. Then, the scattering data at $t = 0$, that is $a(k,0)$, $b(k,0)$, $\tau(k,0)$, and $\rho(k,0)$, can be obtained from the Wronskian relations in Eq. (\ref{eqn:wronskian_relations}) along with the definitions of the transmission and reflection coefficients in Eq. (\ref{eqn:transmisson_and_reflection_coefficients}).

\subsubsection{Direct scattering for the AKNS system}

Similarily, the eigenfunctions $\boldsymbol{\phi}$ and $\boldsymbol{\psi}$ of the AKNS system solve linear integral equations with convergent Neumann series \cite{Ablowitz2011} and the initial scattering data $a(k,0)$, $\overline{a}(k,0)$, $b(k,0)$, and $\overline{b}(k,0)$ is constructed from the Wronskian relations in Eqs. (\ref{eqn:AKNS_wronskian_relations_a}) and (\ref{eqn:AKNS_wronskian_relations_b}).

\subsection{Time evolution of the scattering data}

The scattering data evolve in time according to elementary differential equations.

\subsubsection{Time evolution for the time-independent Schr\"odinger equation}

Following the procedure in cf. Ref. \cite{Ablowitz2011} the time dependence of the scattering data is given by 
\begin{align}
    a(k,t) &= a(k,0), \quad b(k,t) = b(k,0) e^{- 2 i k \gamma(k^2) t},
    \label{eqn:scattering_time_dependence_a_b} \\
    \rho(k,t) &= \rho(k,0) e^{\!- 2 i k \gamma(k^2) t}, ~c_{j}(t) = c_{j}(0) e^{\kappa_{j} \gamma(-\kappa_{j}^2) t} \label{eqn:scattering_time_dependence_rho_c}
\end{align}
where $c_j^2(t)=-i b_j(t)/a_{j}'(t)$ with $a_{j}'(t) \equiv \partial_{k}a(k,t)|_{k = i \kappa_{j}}$ and $\gamma$ comes from the nonlinear evolution equation in Eq. (\ref{eqn:general_evolution_equation_SM}). Notice that $a(k,t)$ is a constant of motion.

\subsubsection{Time evolution for the AKNS system}

For the general evolution equation associated with the AKNS system in Eq. (\ref{eqn:AKNS_general_evolution_equation_SM}), we find the time evolution to be \cite{Ablowitz2011}
\begin{alignat}{2}
    a(k,t) &= a(k,0), &&\overline{a}(k,t) = \overline{a}(k,0), \label{eqn:AKNS_timedep_a}\\
    b(k,t) &= b(k,0) e^{- 2 A_{0}(k) t}, ~  &&\overline{b}(k,t) = \overline{b}(k,0) e^{2 A_{0}(k), t} \label{eqn:AKNS_timedep_b}\\
    C_j(t) &=C_j(0)  e^{- 2 A_{0}(k_j) t}, ~  &&\overline{C}_j(t) = \overline{C}_j(0) e^{2 A_{0}(\overline{k}_j) t}\
    \label{eqn:AKNS_timedep_C}
\end{alignat}
where $C_{j}(t) = b_{j}(t)/a_{j}'(t)$ and $\overline{C}_{j}(t) = \overline{b}_{j}(t)/\overline{a}_{j}'(t)$ with $a_{j}'(t) \equiv \partial_{k}a(k,t)|_{k = i \kappa_{j}}$ and $\overline{a}_{j}'(t) \equiv \partial_{k}\overline{a}(k,t)|_{k = i \overline{\kappa}_{j}}$. Recall that $A_{0}(k)$ is related to the nonlinear evolution equation in Eq. (\ref{eqn:AKNS_general_evolution_equation_SM}).

\subsection{Inverse Scattering}

Inverse scattering is analogous to the inverse Fourier transform, except evaluating an integral on the real line in the case of Fourier transforms is now replaced by solving a linear integral equation in the case of the IST.

\subsubsection{Inverse scattering for the time-independent Schr\"odinger equation}

The inverse scattering and solution $q(x,t)$ of fKdV can be constructed by solving the following Gel'fand-Levitan-Marchenko (GLM)  integral equation for $K(x,y;t)$:
\begin{align}
    K(x,y;t) &+ \tilde{F}(x+y;t) + \!\int_{x}^{\infty} \! ds K(x,s;t) \tilde{F}(s+y;t) = 0, \label{eqn:GLM_equation} \\
    \tilde{F}(x;t) &= \sum_{j = 1}^{J} c_{j}^{2}(t) e^{- \kappa_{j} x} + \frac{1}{2\pi} \int_{-\infty}^{\infty} dk \rho(k,t)e^{i k x}. \notag 
\end{align}
Here $y>x$ and recall that $J$ is the number of zeros of $a$ or, equivalently, the number of solitons in the solution $q$. Here, the time dependence of $\rho(k,t)$ and $c_{j}(t)$ are given in Eq. (\ref{eqn:scattering_time_dependence_rho_c}). Then the solution of the fKdV is obtained from 
\begin{align}
    q(x,t) = 2 \frac{\partial}{\partial x} K(x,x;t). \label{eqn:solution_from_K}
\end{align}

\subsubsection{Inverse scattering for the AKNS system}

The solution of fNLS and the general fractional $q,r$ system can be constructed by solving the following GLM-type integral equations
\begin{align}
    \mathbf{K}(x,y;t) &+ \begin{pmatrix} 1 \\ 0 \end{pmatrix} \overline{F}(x+y;t) \label{1.72b} \\
    &+ \int_{x}^{\infty} ds \, \overline{\mathbf{K}}(x,s;t) \overline{F}(s+y;t) = 0, \notag
\end{align}
\begin{align}
    \overline{\mathbf{K}}(x,y;t) &+ \begin{pmatrix} 0 \\ 1 \end{pmatrix} F(x+y;t) \label{1.72a} \\
    &+ \int_{x}^{\infty} ds \, \mathbf{K}(x,s;t) F(s+y;t)=0 \notag
\end{align}
where
\begin{align}
    F(x;t) &=
    \frac{1}{2\pi} \int_{-\infty}^{\infty} dk \rho(k,t) e^{ik x} - i\sum_{j=1}^{J}C_{j}(t) e^{ik_{j}x},\\
    \overline{F}(x;t) &= \!\frac{1}{2\pi} \! \int_{-\infty}^{\infty} \! dk \overline{\rho}(k,t) e^{\!-ik x} + i\sum_{j=1}^{\overline{J}}\overline{C}_{j}(t)e^{\!-i\overline{k}_{j}x}.
\end{align}
The time dependence of $\rho (k,t)=b(k,t)/a(k)$,  $\overline{\rho}(k,t)=\overline{b}(k,t)/\overline{a}(k)$, $C_j(t)$, and $\overline{C}_{j}(t)$ are given in Eqs. (\ref{eqn:AKNS_timedep_a})-(\ref{eqn:AKNS_timedep_C}).

%
%
The solution of the fractional $q,r$ system is obtained from
\begin{align}
    q(x,t) = -2 K^{(1)}(x,x;t),~ r(x,t) = -2 \overline{K}^{(2)}(x,x;t)  \label{1.75}
\end{align}
where $K^{(n)}$ and $\overline{K}^{(n)}$ for $n=1,2$ denote the $n$th component of the vectors $\mathbf{K}$ and $\overline{\mathbf{K}}$ respectively. If the symmetry $r=\mp q^{\ast }$ holds then the GLM equations have the scalar reduction
\begin{align}
    \overline{F}(x;t)=\mp F^{\ast }(x;t) \notag
\end{align}
and consequently
\begin{align}
    \overline{\mathbf{K}}(x,y;t) = \left(\begin{array}{c} K^{(2)}(x,y;t) \\ \mp K^{(1)}(x,y;t)\end{array}\right)^{\ast}.
\end{align}
In this case 
the inverse problem
reduces to
\begin{align}
    &K^{(1)}(x,y;t) = \pm F^{\ast}(x+y;t) \\
    &\mp \int_{x}^{\infty} ds \int_{x}^{\infty}ds^{\prime} K^{(1)}(x,s^{\prime};t) F(s+s^{\prime};t) F^{\ast}(y+s;t) \notag
\end{align}
and an equation for $K^{(2)}$ which we do not write here. Then the solution to fNLS can be obtained from Eq. \eqref{1.75}. We
also note that when $r= \mp q$ with $q$ real, then $F(x;t)$ and $K^{(1)}(x,y;t)$
are real.

\subsection{Alternative Representation of the fKdV operator}
The operator $\gamma(L^A)$ acting on an arbitrary function $h(x)$ may be represented by the spectral expansion
\begin{align}
 \gamma(L^A) h(x) &= \int_{\Gamma_{\infty}}\! dk\gamma(k^2)\frac{\tau^2(k)}{4 \pi i k}\int_{-\infty}^{\infty}\! dy G(x,y,k) h(y)
 \label{eqn:gamma_L_spectral_expansion_SM}
\end{align}
in terms of the eigenfunctions of the Schr\"odinger scattering problem where time $t$ is suppressed. This expression can be evaluated using contour integration to give a representation of $\gamma(L^A)$ in terms of integrals along the real line and a sum over discrete values along the imaginary axis:
\begin{align}
 \gamma(L^A) h(x) &= \int_{-\infty}^{\infty} dk\,\gamma(k^2)\frac{\tau^2(k)}{4 \pi i k}\,\int_{-\infty}^{\infty} dy \,G_{c}(x,y,k) h(y) \notag \\
 &+ \sum_{j = 1}^{J} \gamma(- \kappa_{j}^2) \int_{-\infty}^{\infty} dy \,G_{d,j}(x,y,k) h(y).
 \label{eqn:gamma_contour_integration}
\end{align}
Here the continuous contribution $G_{c}$ is defined by
\begin{align}
    G_{c}(x,y,k) = \partial_{x} \left(\psi^2(x,k) \varphi^2(y,k) - \varphi^2(x,k) \psi^2(y,k) \right), \label{eqn:G_continuous}
\end{align}
and the discrete contribution, which comes from the poles of $\tau$ at $k_{j} = i \kappa_{j}$, $j = 1,2,...,J$, is given by
\begin{align}
    &G_{d,j}\! = i \eta_{j} \frac{\partial}{\partial x} \left[ \psi^2(y) \psi(x) \partial_{k}\varphi(x)\! -\! \psi^2(y) \varphi(x)\partial_{k}\psi(x) \right]_{k = k_{j}} \notag \\
    &-i \eta_{j} \frac{\partial}{\partial x} \left[ \psi^2(x) \psi(y) \partial_{k}\varphi(y) - \psi^2(x) \varphi(y) \partial_{k}\psi(y) \right]_{k = k_{j}} \label{eqn:G_discrete}
\end{align}
where $\eta_{j} = \frac{b(k_{j})}{\kappa_{j} a'(k_{j})^2}$. Note that we have suppressed $k$ in the above expression, i.e., $\psi(x) = \psi(x,k)$ and $\varphi(x) = \varphi(x,k)$.

\subsection{Explicit form of the fNLS equation}

The set of nonlinear evolution equations which may be associated with the AKNS scattering problem, Eqs. (\ref{eqn:AKNS_scattering_1_SM}-\ref{eqn:AKNS_scattering_2_SM}), is given in Eq. (\ref{eqn:AKNS_general_evolution_equation_SM}).
%
Further, $A_{0}(\mathbf{L}^{A})$ may be represented as
\begin{align}
    A_{0}(\mathbf{L}^{A}) \mathbf{v}(x) \! &= \!\!\sum_{n = 1}^{2}\!\int_{\Gamma_{\infty}^{(n)}}\! \!\! dk  A_{0}(k) f_{n}(k) \!\! \int_{-\infty}^{\infty}\! dy\mathbf{G}_{n}(x,\!y,\!k)\mathbf{v}(y).\label{eqn:A0_L_spectral_expansion_SM}
\end{align}
We obtain fNLS by taking $A_{0}(\mathbf{L}^{A}) = 2 i (\mathbf{L}^{A})^2 |2 \mathbf{L}^{A}|^{\epsilon}$ and $r = \mp q^{*}$ in Eq. (\ref{eqn:AKNS_general_evolution_equation_SM}). If we split off $2 i (\mathbf{L}^{A})^2$ and operate on $\mathbf{u} = \left(r,q\right)^{T}$, we find
\begin{align}
    2 i (\mathbf{L}^{A})^2 \mathbf{u} = \frac{1}{2 i} \begin{pmatrix} \mp q_{xx}^{*} - 2 (q^{*})^2 q \\ q_{xx} \pm 2 q^2 q^{*} \end{pmatrix}.
\end{align}
Then, representing $|2 \mathbf{L}^{A}|^\epsilon$ with the spectral expansion in Eq. (\ref{eqn:A0_L_spectral_expansion_SM}), we have
\begin{align}
    2 i (\mathbf{L}^{A})^2 |2 \mathbf{L}^{A}|^{\epsilon} \mathbf{u} = \sum_{n = 1}^{2} \frac{1}{2 i}\int_{\Gamma_{\infty}^{(n)}} \!\! dk  |2 k|^{\epsilon} f_{n}(k) \\ \times \int_{-\infty}^{\infty} dy \mathbf{G}_{n}(x,y,k) \begin{pmatrix} \mp q_{yy}^{*} - 2 (q^{*})^2 q \\ q_{yy} \pm 2 q^2 q^{*} \end{pmatrix}. \label{eqn:fNLS_operator}
\end{align}
Putting Eq. (\ref{eqn:fNLS_operator}) into Eq. (\ref{eqn:AKNS_general_evolution_equation_SM}), multiplying by $-i$, and taking the second component gives
\begin{align}
    i q_{t} = \sum_{n = 1}^{2} \int_{\Gamma_{\infty}^{(n)}} \!\! dk  |2 k|^{\epsilon} f_{n}(k) \int_{-\infty}^{\infty} dy F_{n}(x,y,k)
\end{align}
where
\begin{align}
    F_{1}(x,y,k) =\, &-\phi_{1}^{2}(x,k) \big[ \psi_{1}^2(y,k) (\mp q_{yy}^{*} - 2 (q^{*})^{2} q ) \\
    &+\psi_{2}^{2}(y,k) (q_{yy} \pm 2 q^2 q^{*}) \big]
\end{align}
and $F_{2}$ is the same, but with $\psi_{n}$ replaced by $\overline{\psi}_{n}$ and $\phi_{n}$ replaced by $\overline{\phi}_{n}$ for $n = 1,2$ where $\overline{\psi}_{n}$ and $\overline{\phi}_{n}$ are related to $\psi_{n}$ and $\phi_{n}$ by equation (\ref{eqn:ANKS_conjugate_relation})
\subsection{Conserved Quantities}

Like their integer counterparts, the fKdV and the fNLS equation also admit an infinite number of conserved quantities. In fact, the derivation of these using IST methods is independent of the exact form of $\gamma$ and $A_{0}$, so their conserved quantities are the same as KdV and NLS; cf. \cite{Ablowitz1}. However, the fluxes associated with these conservation laws corresponding to these conserved quantities are not the same.

\subsection{Direct calculation for the fKdV soliton}

It can be shown that the fractional soliton solution given by
\begin{align}
    q_{K}(x,t) &= 2 \kappa^{2} \text{sech}^{2}{w_{\epsilon}(x,t)} \label{eqn:KdV_one_soliton_SM}
\end{align}
solves the fKdV equation where $w_{\epsilon}(x,t) = \kappa [(x - x_{1}) -(4 \kappa^2)^{1+\epsilon} t]$. This one soliton corresponds to $J = 1$, i.e., one bound state solution of the time-independent Schr\"odinger equation, (\ref{eqn:schrodinger_scattering_SM}), at $k = i \kappa$ and a reflectionless potential $\rho(k,t) = 0$ for real $k$. To show this, we verify that the fKdV equation 
\begin{align}
    q_{t} +\! \int_{\Gamma_{\infty}}\!\! dk |4 k^2|^{1+\epsilon}  \frac{\tau^2(k)}{2 \pi i k}\int_{-\infty}^{\infty} dy\, G(x,y,k) q_{y}\! =\! 0 \label{eqn:fKdV2}
\end{align}
is satisfied when $q = q_{K}$.
For this case, the Schr\"odinger eigenfunctions --- which are found by solving Eq. (\ref{eqn:schrodinger_scattering_SM}) with $q = q_{K}$ --- can be written as
\begin{align}
    \psi(x,k,t) &= v_{1}(x,k,t), \label{eqn:one_soliton_eigenfunction_psi}\\
    \varphi(x,k,t) &= v_{-1}(x,k,t) \label{eqn:one_soliton_eigenfunction_phi}
\end{align}
where
\begin{align}
    v_{\sigma}(x,k,t) = e^{i \sigma k x} \left( \frac{k + \sigma i \kappa \tanh{w_{\epsilon}(x,t)}}{k + i \kappa} \right)
\end{align}
 with $\sigma = \pm 1$. Further, $\partial_{x} q_{K}$ is
\begin{align}
    \partial_{x}q_{K}(x,t) = - 4 \kappa^{3} \text{sech}^2{w_{\epsilon}(x,t)} \tanh{w_{\epsilon}(x,t)}.
\end{align}
To calculate $\gamma(L^A)\partial_{x}q_{K}$, we compute the integrals $I_{c} \equiv \int_{-\infty}^{\infty} dy \, G_{c}(x,y,k) \partial_{y}q_{K}(y,t)$ and $I_{d} \equiv \int_{-\infty}^{\infty} dy \, G_{d,1}(x,y,k) \partial_{y}q_{K}(y,t)$ according to Eqs. (\ref{eqn:gamma_contour_integration})-(\ref{eqn:G_discrete}). These two integral can be written as
\begin{align}
    I_{c} &= \partial_{x} (v_{1})^2 I_{-1}^{(1)} + \partial_{x} (v_{-1})^2 I_{1}^{(1)} |_{k = i \kappa}, \label{eqn:continuous_integral} \\
    I_{d} &= i \eta \partial_{x} \left( v_{1} \partial_{k}v_{-1} - v_{-1} \partial_{k} v_{1}\right) I_{1}^{(1)} |_{k = i \kappa} \label{eqn:discrete_integral} \\
    & - i \eta \partial_{x} (v_{1})^{2} \left( I^{(2)}_{1} - I^{(2)}_{-1} \right) |_{k = i \kappa} \notag
\end{align}
where
\begin{align}
    I_{\sigma}^{(1)} &= \int_{-\infty}^{\infty} dy ~ v_{\sigma}^2(y,k,t) q_{y}(y,t), \label{eqn:I_1} \\
    I_{\sigma}^{(2)} &=  \int_{-\infty}^{\infty} dy ~ v_{\sigma}(y,k,t) \partial_{k} v_{-\sigma}(y,k,t) q_{y}(y,t) \label{eqn:I_2}
\end{align}
and $\eta = b(i \kappa,t)/\kappa (\partial_{k}a(k,t)|_{k = i \kappa})^2$
Using the substitution $w = w_{\epsilon}(y,t)$, the integral $I_{\sigma}^{(1)}$ can be shown to be proportional to $\int_{-\infty}^{\infty} dw ~ f_{\sigma}(w)$ where
\begin{align}
    f_{\sigma}(w) = e^{2 i \sigma k w/\kappa} \left( k + i \sigma \kappa \tanh{w} \right)^2 \text{sech}^2{w} \tanh{w}.
\end{align}
If we consider the rectangular contour with corners at $w = -R$ and $w = R + i \pi$ and define the bottom, top, right, and left contour by $C_{B}$, $C_{T}$, $C_{R}$, and $C_{L}$ respectively, we have
\begin{align}
    \left(\int_{C_{B}} dw + \int_{C_{T}} dw + \int_{C_{R}} dw + \int_{C_{L}} dw \right) f_{\sigma}(w) \label{eqn:f_sigma} \\
    = 2 \pi i \text{Res}\left(f_{\sigma},i \pi/2\right) \notag.
\end{align}
However, the contours along $C_{R}$ and $C_{L}$ vanish as $R\to\infty$ and $C_{B}$ may be written in terms of $C_{T}$ as
\begin{align}
    \int_{C_{T}} dw ~ f_{\sigma}(w) = - e^{- 2 \sigma k \pi/\kappa} \int_{C_{B}} dw ~ f(w).
\end{align}
Because as $R\to\infty$, $\int_{C_{B}} dw f_{\sigma}(w) \to \int_{-\infty}^{\infty} dw f_{\sigma}(w)$, equation (\ref{eqn:f_sigma}) becomes
\begin{align}
    \left(1 - e^{-2 \sigma k \pi/\kappa} \right)\int_{-\infty}^{+\infty} dw f_{\sigma}(w) = 2 \pi i \text{Res}\left(f_{\sigma},i \pi/2\right).
\end{align}
But as the residue of $f_{\sigma}$ vanishes when $\sigma = \pm 1$, $\int_{-\infty}^{+\infty} dw f_{\sigma}(w) = 0$ and, thus, $I_{\sigma}^{(1)} = 0$. Therefore, the continuous contribution to $\gamma(L^A) q_{x}$, $I_{c}$, vanishes. We also see that the first half of the discrete part, $I_{d}$ given by Eq. (\ref{eqn:discrete_integral}), is zero. With these reductions, we can express $\gamma(L^A)\partial_{x}q_{K}$ as
\begin{align}
    \gamma(L^A)\partial_{x}q_{K} = - i \eta \gamma(- \kappa^2) \partial_{x} (v_{1})^{2} \left( I^{(2)}_{1} - I^{(2)}_{-1} \right) |_{k = i \kappa}.
\end{align}
By explicitly evaluating $\eta$ and $\partial_{x} (v_{1})^{2}$, we find that
\begin{align}
    \eta \partial_{x} (v_{1})^2|_{k = i \kappa} &= \frac{\partial_{x}q_{K}}{2 \kappa}.
\end{align}
Then, the integral $I_{\sigma}^{(2)}$ can be evaluated using the fundamental theorem of calculus to give
\begin{align}
    \left( I_{1}^{(2)} - I_{-1}^{(2)} \right) |_{k = i \kappa} = 2 i \kappa.
\end{align}
So, $\gamma(L^A)\partial_{x}q_{K}$ follows as
\begin{align}
    \gamma(L^A) \partial_{x}q_{K} = - (2 \kappa)^{2+2\epsilon} \partial_{x}q_{K}(x,t)
\end{align}
which is exactly $-\partial_{t}q_{K}(x,t)$. Thus, the soliton given in Eq. (\ref{eqn:KdV_one_soliton_SM}) solves the fKdV equation.

\bibliography{refs}
\end{document}